\documentclass[aps,pra,twocolumn,
floatfix,
10pt,english,superscriptaddress,a4paper
]{revtex4-1}

\usepackage[utf8]{inputenc}
\usepackage[T1]{fontenc}
\usepackage{babel}
\usepackage{amsmath}
\usepackage{textcomp}
\usepackage{graphicx}
\usepackage{xcolor}
\usepackage{hyperref}

\hypersetup{
  breaklinks	= true,
  colorlinks	= true,	%Colours links instead of ugly boxes
  urlcolor	= violet,	%Colour for external hyperlinks
  linkcolor	= blue,		%Colour of internal links
  citecolor	= violet	%Colour of citations
}

\newcommand{\ket}[1]{\vert{#1}\rangle}
\newcommand{\bra}[1]{\langle{#1}\vert}
\newcommand{\bracket}[2]{\langle{#1}\vert{#2}\rangle}
\newcommand{\proj}[1]{\ket{#1}\!\bra{#1}}
\newcommand{\mean}[1]{\langle #1 \rangle}
\newcommand{\e}{\mathrm{e}}
\newcommand{\dd}{\mathrm{d}}
\renewcommand{\vec}[1]{{\boldsymbol{#1}}}

\newcommand{\abs}[1]{\lvert#1\rvert}
\begin{document}

%%%%%%%%%%%%%%%%%%%%%%%%%%%%%%%%%%%%%%%%%%%%%%%%%%%%%%%%%%%%%%%%%%%%%%%%%%%%%%%
\title{	Adaptive quantum computation in changing environments
		using projective simulation}
%%%%%%%%%%%%%%%%%%%%%%%%%%%%%%%%%%%%%%%%%%%%%%%%%%%%%%%%%%%%%%%%%%%%%%%%%%%%%%%

\author{M. Tiersch}
\email[Correspondence to ]{markus.tiersch@uibk.ac.at}
\affiliation{Institute for Theoretical Physics,
University of Innsbruck,
Technikerstraße~21A, A-6020 Innsbruck, Austria}
\affiliation{Institute for Quantum Optics and Quantum Information,
Austrian Academy of Sciences,
Technikerstraße~21A, A-6020 Innsbruck, Austria}

\author{E. J. Ganahl}
\affiliation{Institute for Theoretical Physics,
University of Innsbruck,
Technikerstraße~21A, A-6020 Innsbruck, Austria}

\author{H. J. Briegel}
\affiliation{Institute for Theoretical Physics,
University of Innsbruck,
Technikerstraße~21A, A-6020 Innsbruck, Austria}
\affiliation{Institute for Quantum Optics and Quantum Information,
Austrian Academy of Sciences,
Technikerstraße~21A, A-6020 Innsbruck, Austria}

\begin{abstract}
Quantum information processing devices need to be robust and stable against external noise and internal imperfections to ensure correct operation. In a setting of measurement-based quantum computation, we explore how an intelligent agent endowed with a projective simulator can act as controller to adapt measurement directions to an external stray field of unknown magnitude in a fixed direction.
We assess the agent's learning behavior in static and time-varying fields and explore composition strategies in the projective simulator to improve the agent's performance.
We demonstrate the applicability by correcting for stray fields in a measurement-based algorithm for Grover's search. Thereby, we lay out a path for adaptive controllers based on intelligent agents for quantum information tasks.
\end{abstract}

\maketitle

%%%%%%%%%%%%%%%%%%%%%%%%%%%%%%%%%%%%%%%%%%%%%%%%%%%%%%%%%%%%%%%%%%%%%%%%%%%%%%%

%======================================================================
\section*{Introduction}
\label{sec:introduction}
%======================================================================

When building devices for quantum information processing one has to take changing environment conditions and device imperfections into account. It is therefore necessary to include adaptive mechanisms that characterize and calibrate the device from within. Furthermore, it is desirable for these devices to obtain a certain degree of autonomy in maintaining their functional state despite detrimental environment influences, in particular, when they are assembled to a larger quantum information processing infrastructure.
In the attempt to miniaturize current implementations of quantum devices, we will reach the point where these devices will be of microscopic scale and require short reaction times. For such microscopic systems we can no longer assume that their internal controllers are full-fledged universal computers that can carry out arbitrary programs. Instead, controllers will be small physical systems that are specialized for their respective purpose with a program that emerges from the controller's analog dynamics.

In this paper we explore the applicability of a controller in form of an intelligent learning agent that has access to a \emph{projective simulator}~\cite{PS,PScompreh,Melnikov14,Melnikov15}. Within this agent framework, the aim is to demonstrate adaptive calibration and compensation strategies against stray external fields when carrying out quantum information tasks. The agent shall thereby implement a simple form of adaptive error avoidance and implicit parameter estimation.

Algorithms from machine learning have been used to find strategies for parameter estimation, and  optimal strategies for parameter estimation are known for specific cases, see e.g.~\cite{Sanders10,Sanders11,Sanders13,Josh,Granade,Hayes}. Here, however, we focus on strategies that arise naturally from the adaptive dynamics of the underlying physical system, for which we choose a projective simulator.
The projective simulator is a platform that has been proposed as a physical model for reinforcement learning~\cite{SuttonBarto,RusselNorvig}, and it effectively reproduces input--output--reward correlations from an internal adaptive stochastic process.
With the restriction to this particular system, one cannot hope for the best possible strategy to emerge while keeping the rules governing the dynamics reasonably simple and computational overhead low. Both requirements are necessary to allow for an actual physical realization.
As an additional feature, the projective simulator offers a natural route to quantization as indicated in~\cite{PS} and thereby a way to intelligent agents that benefit from internal quantum dynamics, as demonstrated in the reflective quantum projective simulator~\cite{rPS,rPSions}. Agent quantization is not explored further in the present work as we focus on the application of a classical agent to quantum information processing first.
For recent comprehensive reviews in the domain of quantum physics and artificial intelligence or machine learning see \cite{WittekBook} and \cite{Petruccione}.

As illustration of our method of adaptive quantum information processing we study Grover's quantum search algorithm \cite{GroverProc,Grover} in the paradigm of measurement-based quantum computation. Grover's algorithm provides a fast way to find a marked item in an unsorted database with $N$ elements. In particular, it provides a quadratic speed-up with $O\big(\sqrt{N}\big)$ database look-ups over a search by means of a classical computer with $O(N)$ look-ups. First proof-of-principle implementations of Grover's algorithm with nuclear magnetic resonance techniques \cite{GroverNMRChuang,GroverNMRJones} and entangled photons \cite{GroverPhotons} employed the circuit model of quantum computation, where individual unitary quantum logic gates are applied to a register of qubits to process information.

Measurement-based quantum computation (MBQC) \cite{MBQC,MBQCbible} is a different paradigm of quantum computation, where the computation is carried out by measuring single qubits of an initially highly entangled resource state~\cite{ClusterStates}. The first experimental demonstration of MBQC in a system of entangled photons~\cite{GroverPhotonsMBQC,PhotonMBQC} (and with trapped ions \cite{IonMBQC}) also demonstrated the Grover algorithm in its smallest realization with a database of 4 entries (2-qubits) by using a 4-qubit cluster state as computation resource.

As preparation for the full measurement-based algorithm we first study a basic setting. We situate a quantum system, a single qubit, in an unknown external magnetic field. An artificial agent, the controller, is endowed with a projective simulator and the ability to measure the quantum system and thereby prepare quantum states. We hardwire the learning process, i.e., the update rule in the reinforcement learning process of the projective simulator, such that the agent effectively carries out the following tasks:
(i) Adapt measurement directions to changes of the external magnetic field, and dynamically improve the sensing resolution.
(ii) Learn to adapt simultaneously for multiple measurement directions needed for general MBQC-algorithms in a feedback scheme.
(iii) Carry out a quantum information task, the Grover algorithm~\cite{GroverProc,Grover} in the setting of measurement-based quantum computation, with unknown stray magnetic fields.
This provides a completely worked-out example, starting from the physical system that generates the actions of an adaptive ``intelligent'' agent, here a projective simulator, to a controller tailored to a specific quantum information task, e.g.\ measurement-based Grover's search algorithm.

%======================================================================
\section*{Results}
\label{sec:results}
%======================================================================

First, we describe an approach that allows the projective simulator to effectively obtain a notion of the strength of an external magnetic field and hence carry out a primitive form of parameter estimation. However, there is a conceptual difference between our approach and parameter estimation. After the agent has learned, the information on the strength of the magnetic field will \emph{not} be available as a number that the agent gives as an output. Instead, this information is only indirectly incorporated into the dynamics and decision patterns of the agent, and it can be exploited to \emph{do} certain things that are adapted to the external field.
Therefore, we will analyze the learning process of the agent from two different perspectives: From an \emph{operational} perspective we characterize how well the agent adapts its actions to the external field, and from an \emph{informational} perspective we quantify how much of the information about the external field is really contained in the parameters that define the dynamics of the agent.

We start with a detailed description of the setting, that is, of the agent and its interaction with the measurement apparatus, the dynamics of the projective simulator, and an analysis of the learning process.

%----------------------------------------------------------------------
\subsection*{Agent and Projective Simulator Dynamics}
\label{sec:agent_and_PS_dynamics}
%----------------------------------------------------------------------

%++++++++++++++++++++++++++++++++++++++++++++++++++++++++++++++++++++++
\begin{figure}[tbp]
	\includegraphics[width=\linewidth]{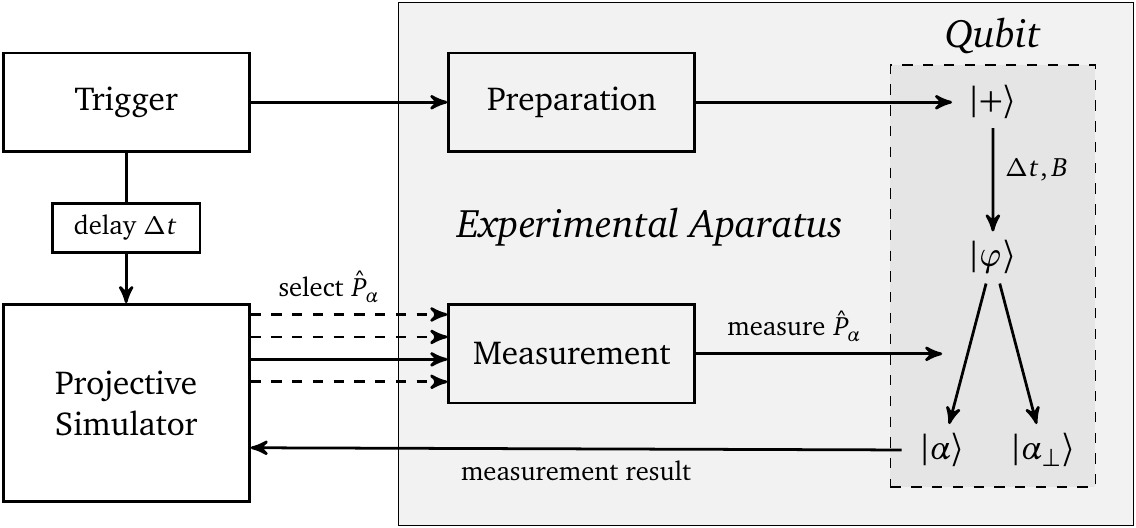}
	\caption{Setup of the agent that is endowed with a projective simulator and operates a measurement apparatus to estimate the state of a single qubit.}
	\label{fig:agentSetup}
\end{figure}
%++++++++++++++++++++++++++++++++++++++++++++++++++++++++++++++++++++++

In the present setting the magnetic field direction is promised to be fixed along the $z$-axis, and the agent needs to estimate its strength $B$. The following steps are visualized in Figure~\ref{fig:agentSetup}. The agent starts by preparing a single qubit in the state $\ket{+}=(\ket{0}+\ket{1})/\sqrt{2}$, which in the presence of the field evolves according to the Hamiltonian $H=\hbar \omega \sigma_z/2$, where the frequency $\omega$ is proportional to the magnetic field strength $B$. After some fixed time interval $\Delta t$, the initial state has evolved into
\begin{equation}
	\label{eq:state_phi}
	\ket{\varphi} = \frac{\ket{0}+\e^{i\varphi}\ket{1}}{\sqrt{2}},
\end{equation}
up to a global phase, with $\varphi=\omega \Delta t$. Estimating the field strength $B$ amounts to estimating $\ket{\varphi}$ and obtaining information about the angle $\varphi$ between this state and the initial state in the equatorial plane of the Bloch sphere.
In a linear optics setup~\cite{KLM}, the unknown angle $\varphi$ would correspond to an unknown phase shifter in the beam line.

The agent measures the qubit in the unknown state $\ket{\varphi}$ in various directions and incorporates the measurement outcomes to change its choice of measurement directions.
The measurements applied by the agent are in general described by POVMs~\cite{Helstrom}. For simplicity, we will restrict our analysis to projective measurements. We shall comment on the general case at the end of the paper.

The challenge is to effectively realize a probability distribution for the unknown angle $\varphi$ without explicitly performing computations and analyzing the measurement data. Rather it should emerge dynamically as the result of a feedback loop by reinforcing certain actions on the quantum system.
Therefore, we choose an approach where the internals of the agent are wired such that it tries to optimize the direction of a measurement. In the optimal case $\ket{\varphi}$ is the $+1$ eigenstate of this measurement.
Qubit observables whose eigenstates with eigenvalues $\pm1$ lie in the equator of the Bloch sphere are given by
\begin{equation} \label{eq:observable}
	\hat{O}_\alpha = \proj{\alpha} - \proj{\alpha+\pi},
\end{equation}
where $\ket{\alpha}$ is of the form \eqref{eq:state_phi} with angle $\alpha$. Both eigenstates lie on opposite sides of the equator. The probability to obtain the measurement outcome $\pm1$ is
\begin{equation} \label{eq:prob_result}
	p(\pm1\vert\varphi,\alpha)=\frac{1\pm\cos(\varphi-\alpha)}{2},
\end{equation}
that is, the closer the angles $\alpha$ and $\varphi$ the higher is the probability to obtain the $+1$ measurement outcome. To simplify notation we often consider the projector onto the $+1$ eigenstate
\begin{equation}
	\hat{P}_\alpha = \proj{\alpha}
\end{equation}
instead of the observable $\hat{O}_\alpha$, and measurements of the projector with outcomes 1 and 0. For qubits, measuring $\hat{P}_\alpha$ gives the same statistics of measurement outcomes and resulting states as measuring the observable $\hat{O}_\alpha$ because there is a unique state orthogonal to $\ket{\alpha}$.

\begin{figure}[tbp]
	\includegraphics{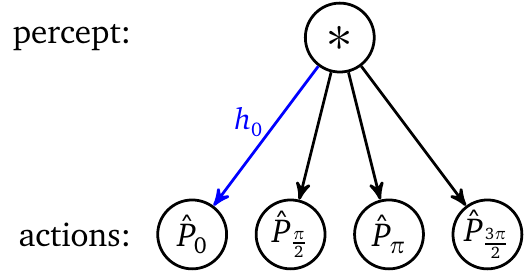}
	\caption{Clip network of the projective simulator. The stochastic process is initialized by an excitation at the $*$-clip, which then undergoes random walk dynamics according to the weights of the links. The action clip where the excitation arrives determines the measurement direction.}
	\label{fig:network_1to4}
\end{figure}

The projective simulator inside the agent employs an adaptive stochastic process that is modeled by a random walk of an excitation in a network of so-called ``clips''~\cite{PS}. For now the clip network takes the form of a directed weighted graph depicted in Figure~\ref{fig:network_1to4}. The random walk starts at the only ``percept clip'', which is excited by an internal trigger of the agent with a time interval $\Delta t$ after the qubit has been prepared (cf.\ Figure~\ref{fig:agentSetup}). The excitation propagates in the network according to the weights of the links that connect the percept clip to the action clips. Once the excitation reaches an action clip, the corresponding action is performed and the process inside the projective simulator is finished. A single action is a measurement of a certain $\hat{P}_\alpha$ at the qubit. If the measurement outcome is $+1$ it is fed back as reward to the agent to re-enforce and strengthen the link between the percept clip and the last action clip. The process is repeated for the next measurement. The probabilities to select certain measurements, however, change as a result of previous measurement outcomes. This makes measurements with angles $\alpha$ closer to $\varphi$ more likely. These probabilities in effect represent a coarse-grained, discrete probability distribution over angles $\varphi$.

%++++++++++++++++++++++++++++++++++++++++++++++++++++++++++++++++++++++
%++++++++++++++++++++++++++++++++++++++++++++++++++++++++++++++++++++++

%++++++++++++++++++++++++++++++++++++++++++++++++++++++++++++++++++++++
\begin{figure*}[tbp]
	\includegraphics[width=\linewidth]{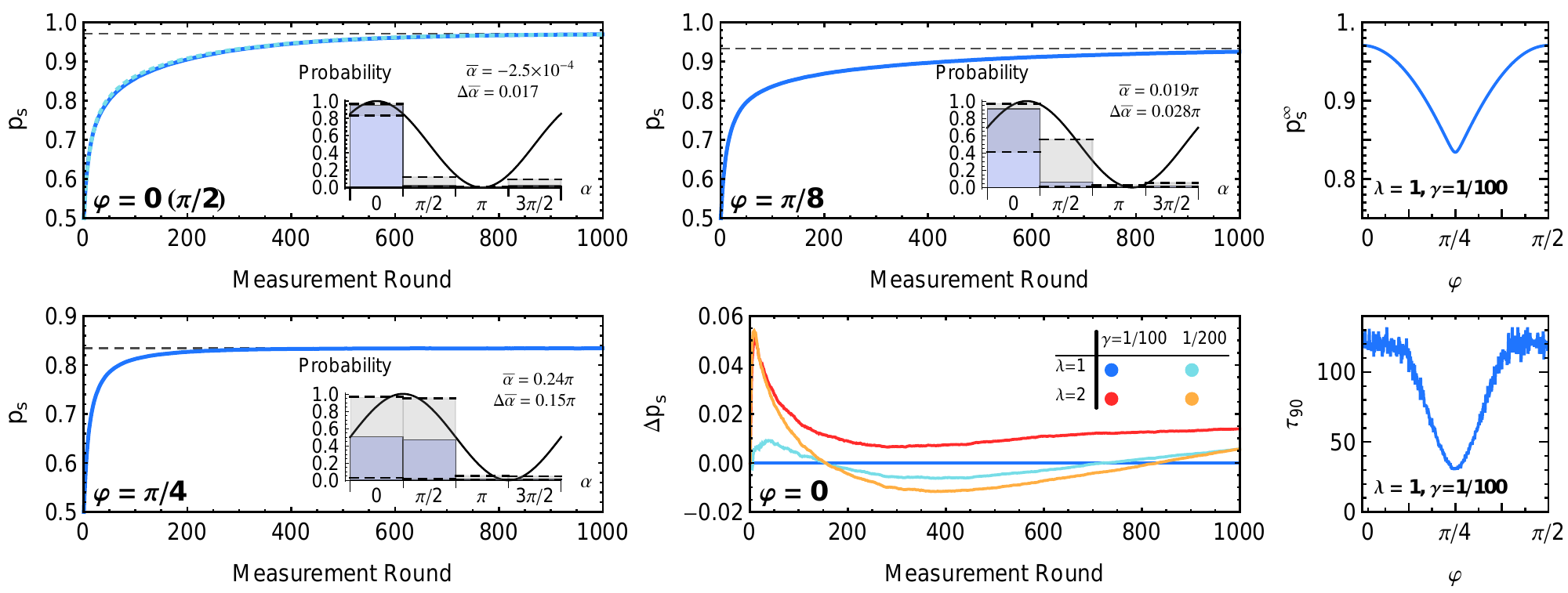}
	\caption{\textbf{Learning curves.}
	\textbf{Left:} The success probability $p_s$ as a function of number of measurement rounds on the qubit in state $\ket{\varphi}$ is shown for four angles as averaged over an ensemble of $N=1000$ agents running in parallel ($\lambda=1$, $\gamma=1/100$). The time scale of learning can be rescaled by increasing both $\lambda$ and $\gamma$. Dashed lines give analytical approximations of the asymptotic values (see text). The insets show the transition probabilities of the average agent after 1000 learning steps, $\bar{h}_\alpha/\sum_{\alpha^\prime}\bar{h}_{\alpha^\prime}$, where $\bar{h}_\alpha$ is the ensemble average of the weight $h_\alpha$, together with the minimal and maximal probabilities obtained in the ensemble (as error bars), and the analytical curve of $p(+1\vert\varphi,\alpha)$. From the transition/measurement probabilities of a single agent $j$ we infer \cite{Fisher,Mardia} its mean angle $\mean{\alpha}_j=\arg \sum_\alpha p_\alpha \e^{i\alpha}$. The ``vector sum'' of the mean angles of the $N$ individual agents is the complex number $\vec{r}=\sum_j \e^{i \mean{\alpha}_j}/N$, which determines the ensemble average of the mean angle $\bar{\alpha}=\arg\vec{r}$ and its circular standard deviation $\Delta \bar{\alpha}=(-2\ln\abs{\vec{r}})^{1/2}$.
	\textbf{Bottom middle:} A higher reward scaling $\lambda$ and lower damping rate $\gamma$ give a faster initial learning and a higher asymptote, with a slower final convergence for the latter. Curves show the differences of all $p_s$ with the reference case $\lambda=1$, $\gamma=1/100$.
	\textbf{Top right:} Asymptotic success probability $p_s^\infty$ (analytical approximation) as a function of $\varphi$ for 4 projectors. The curve is $\pi/2$-periodic.
	\textbf{Bottom right:} Learning time $\tau_{90}$ for the ensemble of agents to reach 90\,\% of $p_s^\infty$. Data is the average of the learning times of 1000 agents.}
	\label{fig:learning_single_phi}
\end{figure*}
%++++++++++++++++++++++++++++++++++++++++++++++++++++++++++++++++++++++

In detail, each link in the clip network carries a weight $h_\alpha$. The probability to jump from the percept clip ``$*$'' to the action clip corresponding to $\hat{P}_\alpha$ is given by the normalized weight of all edges from the percept clip, that is,
\begin{equation}
	p_\alpha = \frac{h_\alpha}{\sum_{\alpha^\prime} h_{\alpha^\prime}}.
\end{equation}
At the beginning of the learning process all weights are initialized with $h_\alpha(0)=1$.
After the measurement of $\hat{P}_\alpha$ in the $n$-th round, the measurement outcome (0 or 1) is rescaled by a factor $\lambda$ and fed back into the projective simulator as a reward $\lambda_n$ to the transition with weight $h_\alpha$. Regardless of whether or not a transition has been taken, all weights are damped by a small amount with rate $\gamma$. After the $n$-th round, in which $\alpha$ was the measurement angle, all weights are changed according to the following update rule:
\begin{equation} \label{eq:updateRule}
	h_{\alpha^\prime}(n+1) = h_{\alpha^\prime}(n) - \gamma\Big( h_{\alpha^\prime}(n) - 1\Big) + \delta_{\alpha \alpha^\prime} \lambda_n.
\end{equation}
As a result, the projective simulator converges to a state (set of $h$-values) that increases the chances of obtaining $+1$ measurement outcomes and thereby increases the probability to measure in directions close to $\varphi$.
From the perspective of the projective simulator only an outcome $+1$ denotes success because the action that led to this outcome will be reinforced. This ``subjective'' success probability is
\begin{equation}
	p_s \equiv p(+1\vert\varphi) = \sum_\alpha p(+1\vert\varphi,\alpha) \, p_\alpha.
\end{equation}
An action that leads to a reward (measurement result $+1$) is also the correct action from an operational point of view.
The transition probabilities $p_\alpha$ provide an internal representation of a discretized probability distribution for the angle $\varphi$. The change of $p_s$ as a function of the number of rounds (measurements on the quibt) is depicted in Figure~\ref{fig:learning_single_phi} for several examples of $\varphi$. The results in Figure~\ref{fig:learning_single_phi} show that the agents learns to obtain rewards more often and thus obtains information about the state $\ket{\varphi}$ and thereby about~$B$.

%----------------------------------------------------------------------
\subsection*{Learning Curve Analysis}
\label{sec:learning_curve_analysis}
%----------------------------------------------------------------------

In our example we start with 4 projectors at angles every $\pi/2$, which corresponds to the projectors onto the eigenstates of the observables that are given by the Pauli matrices $\hat{\sigma}_x$ and $\hat{\sigma}_y$.
If $\varphi=0$, measurements of $\hat{P}_0$ will always give outcome $+1$ and hence be rewarded. The two adjacent projectors at $\alpha=\pi/2$ and $3\pi/2$ are rewarded in half of the measurements, and measurements in the direction $\alpha=\pi$ are never rewarded. In this situation the projective simulator builds a strong link to $\hat{P}_0$, somewhat less strong links to $\hat{P}_{\pi/2}$ and $\hat{P}_{3\pi/2}$ and leaves the link for $\hat{P}_\pi$ at its initial value. The coarse-grained discrete probability distribution for $\varphi$ is consequently peaked at $\varphi=0$ and---within statistical fluctuations---symmetric around this direction (Figure~\ref{fig:learning_single_phi} top left inset).
If $\varphi$ is between two of the projectors, say $\varphi\approx\pi/4$, measurements of $\hat{P}_0$ and $\hat{P}_{\pi/2}$ will only be rewarded with only $85\,\%$ probability, and measurements of the opposite projectors with $15\,\%$ probability. The distribution of measurement probabilities will also be symmetric around the direction $\pi/4$ but less pronounced as shown by a broader distribution in Figure~\ref{fig:learning_single_phi} (bottom left inset).
A broad distribution for measuring in the direction $\alpha$ results in a lower success probability for angles that have a large distance to all projectors, e.g., $\alpha=\pi/4$.

At this point a smaller damping rate $\gamma$ and a larger multiplier of the rewards $\lambda$ both lead to a larger value of rewarded transitions in the steady-state and hence to a larger success probability and a probability distribution that is more peaked. At the same time increasing both $\lambda$ and $\gamma$ speeds-up the learning process leading to learning curves with a steeper initial rise. Note, however, that extremal cases with too large rewards or too weak damping favor situations in which the agent prefers actions that just by luck led to a reward in the past although they are not highly rewarded on average. Un-learning such an initial ``misunderstanding'' and building a probability distribution that reflects the actual probabilities of being rewarded may take a long time. This aspect leads to larger fluctuations in the success probability of an ensemble of agents and a slower final convergence.

\textit{Asymptotic Success Probability.}---%
For the asymptotes of the success probability we can find a first-order approximation by assuming a steady state of the transition probabilities $p_\alpha^\infty$ and the respective $h$-values. The resulting steady-state success probability is $p_s^\infty=\sum_\alpha p(+1\vert\varphi,\alpha)p_\alpha^\infty$.
When coarse-graining over many measurements the time average of the reward for each action is given by $\lambda p(+1\vert\varphi,\alpha) p_\alpha^\infty$, and the steady-state probability to measure in direction $\alpha$ is $p_\alpha^\infty=h_\alpha^\infty/\sum_{\alpha^\prime}h_{\alpha^\prime}^\infty$. With these assumptions the update rule \eqref{eq:updateRule} turns into a set of coupled equations for the steady-state values $h_\alpha^\infty$,
\begin{equation}
	(h_\alpha^\infty - 1)\sum_{\alpha^\prime} h_{\alpha^\prime}^\infty = \frac{\lambda}{\gamma} \abs{\bracket{\varphi}{\alpha}}^2 h_\alpha^\infty
\end{equation}
in which the loss terms given by the damping $\gamma$ and the gain terms given by the time-averaged reward are in equilibrium. This set of nonlinear equations can be solved numerically and yields a very good approximation for the ensemble average as seen in Figure~\ref{fig:learning_single_phi}.
The asymptotic value obtained in this approximation only depends on the ratio $\lambda/\gamma$.

%++++++++++++++++++++++++++++++++++++++++++++++++++++++++++++++++++++++
\begin{figure}[tbp]
	\includegraphics[width=\linewidth]{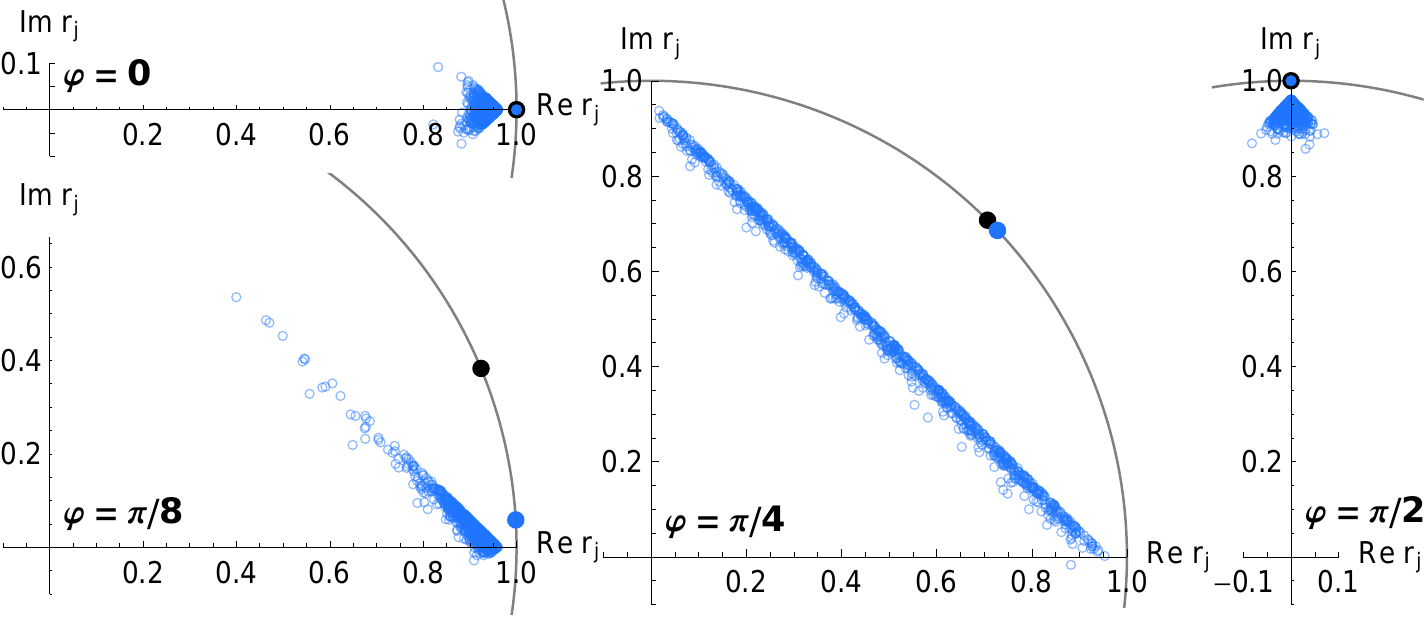}
	\caption{States of each agent in an ensemble of 1000 agents with four available measurement directions after 1000 measurements for $\varphi=0, \pi/8, \pi/4, \pi/2$: States are summarized as the position of the resulting vectors $\vec{r}_j=\sum_\alpha p_\alpha \e^{i\alpha}$ for each agent $j$ in the complex plane (equatorial plane). The degeneracy with respect to the expected reward for $\varphi=\pi/4$ provides a manifold of equally successful states, which is populated by the ensemble. Blue dot on the unit circle gives the angle of the ensemble average, the black dot is the angle $\varphi$.}
	\label{fig:EnsembleAverage}
\end{figure}
%++++++++++++++++++++++++++++++++++++++++++++++++++++++++++++++++++++++

%++++++++++++++++++++++++++++++++++++++++++++++++++++++++++++++++++++++
\begin{figure}[tbp]
	\centering
	\includegraphics[width=\linewidth]{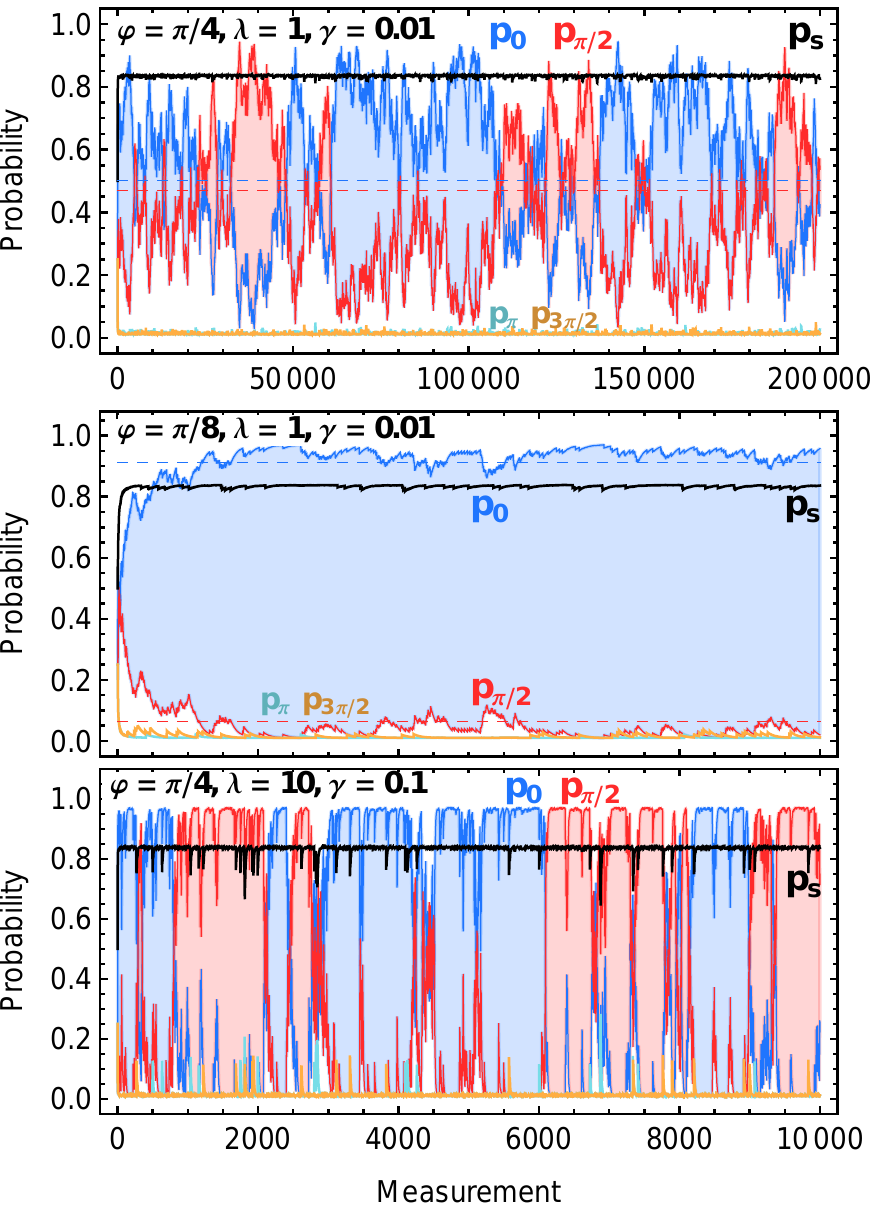}
	\caption{Long-term evolution of the state of a single agent in terms of success probability and the individual probabilities to do one of the four actions. For $\varphi=\pi/4$ (left) the action pair $\hat{P}_0$ and $\hat{P}_{\pi/2}$ is degenerate with respect to the expected reward. For $\varphi=\pi/8$ (middle), i.e., not exactly between two projectors, the agent measures more often into the direction $\alpha=0$. Fluctuations in the measurement probabilities do not necessarily show in the success probability. For comparison, the ensemble averages of 1000 agents after 1000 measurements are given as dashed lines. Larger rewards $\lambda$ and damping $\gamma$ (both rescaled by a factor 10) decrease the timescale of the fluctuations while maintaining approximately the same time average (right). The agent jumps between different preferred action and stays for extended times.}
	\label{fig:TimeAverage}
\end{figure}
%++++++++++++++++++++++++++++++++++++++++++++++++++++++++++++++++++++++

\textit{Time Average vs.\ Ensemble Average.}---%
The fluctuations of the probabilities to choose certain actions (see insets in Figure~\ref{fig:learning_single_phi}) show that even after 1000 iterations of the update rule \eqref{eq:updateRule} not all agents have converged to a single state (Figure~\ref{fig:EnsembleAverage}). Many steady states occur if there is a whole manifold that is rewarded equally, that is, when two or more actions have the same expected reward. For example, for $\varphi=\pi/4$ both actions $\hat{P}_0$ and $\hat{P}_{\pi/2}$ have equal chances of being rewarded and thus the there is no preference of either action as long as one of them is carried out. Actions that have the same expected reward span a subspace for which the sum of the probabilities for doing these actions is approximately constant in the steady state, however, their relative ratio is not. The states of the whole ensemble of agents fills this degenerate subspace of action probabilities. The ensemble average yields an approximation of $\varphi$.

Although the ensemble has learned, i.e., the success probability has converged, the dynamics of each individual is not necessarily converged to a single state where it remains. In the course of time the state of a single agent explores the whole degenerate reward manifold while keeping the success probability constant as we numerically illustrate in Figure~\ref{fig:TimeAverage}. We find that the time average of a single agent for long times equals the ensemble average because the state of the single agent assumes all the different steady states that an ensemble produced after short time as in Figure~\ref{fig:EnsembleAverage}.
Hence, to obtain an ensemble average a snapshot after a relatively short time is sufficient. However, if there is a degenerate space in the reward scheme, the state of a single agent at a fixed time gives only an imprecise estimate of $\varphi$, and even its time average does when considered only for a short time.
A larger damping parameter $\gamma$ and higher rewards $\lambda$ facilitate a faster exploration of the degenerate reward manifold and thus provide a better time average for a single agent for shorter times.
For $\varphi=\pi/4$ in Figure~\ref{fig:TimeAverage}(right), the agent selects either $\alpha=0$ or $\pi/2$ for an extended time and then suddenly switches between these equally rewarded choices. This jumping behavior occurs for large reward and damping, whereas for smaller values (left) also equal probabilities occur for longer durations.

%----------------------------------------------------------------------
\subsection*{Comparison to State Tomography and Bayesian Analysis}
\label{sec:bayesian_analysis}
%----------------------------------------------------------------------

The way that the agent uses the rewards to change its actions to do measurements more often along angles that are close to $\varphi$, is a way of representing information about $\varphi$. We regard this probability distribution of actions along the discrete set of angles as a probability distribution of $\varphi$ \cite{Fisher,Mardia}, and compare it to standard computational analysis procedures employed in state and parameter estimation. By its actions and the returned rewards the agent effectively samples the reward distribution $p(+1\vert\varphi,\alpha)$. The same data, namely the measurement direction and outcome, however, can also be used in a Bayesian update rule to explicitly build a probability distribution $p(\varphi)$, or the data can be used to reconstruct the state $\ket{\varphi}$ via state tomography. We compare the angular distribution of actions that the agent maintains to the angular distribution that a Bayesian update would produce, and also to a simple state tomography by estimating expectation values from the same measurement data.

A simple form of state tomography can be done by calculating the expectation values $\mean{\hat{\sigma}_x}$ and $\mean{\hat{\sigma}_y}$ from the measurement results of the four projective measurements. Together with the initial assumption $\mean{\hat{\sigma}_z}=0$, these expectation values give an approximation of the state's Bloch vector. Our four measurement directions $\alpha=0, \pi/2, \pi, 3\pi/2$ give the same measurements as the Pauli matrices with expectation values
\begin{align}
	\mean{\hat{\sigma}_x}
		&\equiv \mean{\hat{O}_0} \phantom{_{/2}} = -\mean{\hat{O}_\pi}, \\
	\mean{\hat{\sigma}_y}
		&\equiv \mean{\hat{O}_{\pi/2}} = -\mean{\hat{O}_{3\pi/2}},
\end{align}
where expectation values of the observables can be related to those of the projectors by
\begin{equation}
	\mean{\hat{O}_\alpha}=2\mean{\hat{P}_\alpha}-1.
\end{equation}
For a total of $M=\sum_\alpha M_\alpha$ measurements, of which $M_\alpha$ are done in direction $\alpha$, with individual measurement outcomes $r_{\alpha,m}=\pm1$ for observable $\hat{O}_\alpha$, the expectation values can be approximated with the mean
\begin{align}
	\mean{\hat{\sigma}_x}
		&\approx \frac{ \sum_{m=1}^{M_0} r_{0,m} - \sum_{m=1}^{M_\pi} r_{\pi,m} }{M_0 + M_\pi},\\
	\mean{\hat{\sigma}_y}
		&\approx \frac{ \sum_{m=1}^{M_{\pi/2}} r_{{\pi/2},m} - \sum_{m=1}^{M_{3\pi/2}} r_{{3\pi/2},m} }{M_{\pi/2} + M_{3\pi/2}}.
\end{align}
The resulting Bloch vector with coordinates $\big(\mean{\hat{\sigma}_x}, \mean{\hat{\sigma}_y}, 0\big)$ provides an angle with the $x$-axis and thereby an estimate of~$\varphi$.

In a Bayesian analysis, we update an initially flat prior distribution $p(\varphi)=1/(2\pi)$ with the information obtained from each measurement. After each measurement, the distribution is updated with result $r_m\in\{-1,+1\}$ for measurement in direction $\alpha_m$, e.g., for the first update
\begin{equation}
	p(\varphi\vert r_1) = \frac{p(r_1\vert\varphi)p(\varphi)}{p(r_1)},
\end{equation}
where we include the knowledge of quantum mechanics and the statistics of measurement outcomes for the underlying system with $p(r_1\vert\varphi)$ given by \eqref{eq:prob_result}. After $M$ measurements the resulting probability distribution is
\begin{equation}
	p(\varphi\vert r_1,\dotsc,r_M) = \frac{1}{\mathcal{N}} \prod_{m=1}^{M}	\Big( 1+r_m \cos(\varphi-\alpha_m) \Big),
\end{equation}
with normalization $\mathcal{N}$.
For an efficient update and a compact representation of the conditioned probability distribution we expand it in a Fourier series, which has at most $M$ higher harmonics, and construct an recursive update rule for the expansion coefficients following the approach in \cite{Josh} for parameter estimation with a single fixed observable but variable time delays. For our choice of measurement directions, with $\alpha$ being a multiple of $\pi/2$, the Fourier expansion generally contains $\sin$ and $\cos$ terms. The recursive update rules for the expansion coefficients are given in the Appendix.

%++++++++++++++++++++++++++++++++++++++++++++++++++++++++++++++++++++++
\begin{figure}[tbp]
	\centering
	\includegraphics[width=\linewidth]{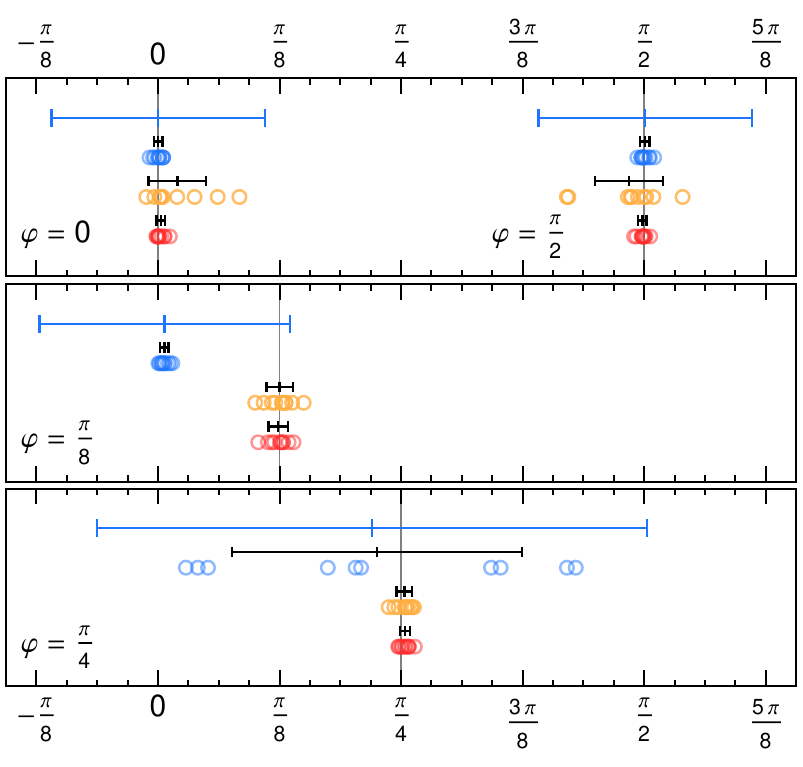}
	\caption{Comparison of angular probability distributions obtained from the projective simulator (blue, top), state tomography by estimation of expectation values (orange, middle), and Bayesian updating (red, bottom) for various $\varphi$. The data points represent the measurement data of 10 agents from 1500 measurements each, and they give an estimate of the angle $\varphi$ as the mean of the 10 distributions. The 10 data points are supplemented by a black error bar, which indicates their circular mean and circular standard deviation. For the projective simulator, the blue error bar indicates the circular mean and circular standard deviation of the discrete probability distribution over the 4 actions, after averaging over all 10 agents.
	For most examples of $\varphi$ the projective simulators generate a distribution of the mean angles that coincide with $\varphi$ except for $\varphi=\pi/8$, where similar as in Figure~\ref{fig:EnsembleAverage} a bias towards the nearest available projector ($\alpha=0$) occurs.
	}
	\label{fig:distributionComparison}
\end{figure}
%++++++++++++++++++++++++++++++++++++++++++++++++++++++++++++++++++++++

To compare the estimates of $\varphi$ by these three approaches, we fix the angle $\varphi$, and let 10 agents with a projective simulator do 1500 measurements each and according to the dynamics arising from using the projective simulator. After these 1500 measurement each agent has built a probability distribution of actions $p_\alpha$, we take the mean of each distribution as the estimate of $\varphi$. Figure~\ref{fig:distributionComparison} shows these estimates as the blue data points. Because the probability distributions for $\varphi$ that we obtain from the $p_\alpha$ have support only on 4 angles, which are uniformly and discretely spaced on the circle, each distribution has a large variance. We average the distributions of all 10 agents and give the mean and circular standard deviation of the resulting distribution as the blue error bar in Figure~\ref{fig:distributionComparison} for comparison. Clearly, when $\varphi$ is close to one of the possible choices of $\alpha$, the projective simulator captures $\varphi$ accurately, but for values of $\varphi=\pi/8$ or $\pi/4$ the estimates are biased towards one of the $\alpha$ as in Figure~\ref{fig:EnsembleAverage}. For $\varphi=\pi/4$ the angular means are widely spread, and their distribution has a large variance, which is reminiscent of the distributions given in the insets in Figure~\ref{fig:learning_single_phi} and the distribution of means of a large ensemble in Figure~\ref{fig:EnsembleAverage}.

The estimates for $\varphi$ obtained from the expectation values of the Pauli matrices, i.e., the simple state tomography, are calculated from the same measurement record for each agent and are given by the orange data points in Figure~\ref{fig:distributionComparison}. For the Bayesian update scheme, we construct the conditional probability distributions for $\varphi$, again from the same measurement record that the projective simulator generated. All of the resulting distributions assume an approximate Gaussian shape with a narrow peak ($\sigma\approx\pi/100$). The means are given as red data points in Figure~\ref{fig:distributionComparison}. Both approaches can estimate $\varphi$ correctly within the error bars. Surprisingly, for $\varphi$ along one of the $\alpha$, the estimates from the expectation values spread more than in the other two approaches. The reason is that in these cases the projective simulator samples most of measurements along a single direction and only few for the other observable, which causes a rather large uncertainty in one of the coordinates.

Although state tomography and Bayesian estimation perform generally equally good or better than the projective simulator, the big conceptual difference between these approaches is that very little knowledge of quantum physics and measurement statistics is build into the projective simulator. The projective simulator does not assume that the rewards originate from measurement probabilities of a quantum state and, therefore, it is ``model free''. The update rule causes a learning dynamics that drive the agent to measure more often into directions that give a +1 measurement outcome and thereby implicitly align measurement directions with $\varphi$. Even when no optimal measurement direction is available the agent learns how to deal with a system such that reward is most likely to occur.
In principle, it could even adapt to artificial situations, where measurements along the $x$-axis always give a $+1$ outcome and measurements along $y$ always give the outcome $-1$, something which cannot be explained by measuring a qubit in a defined fixed state.
Therefore, it is not surprising that methods that make use of additional information, namely measurement probabilities predicted by quantum physics, can extract more information about $\varphi$ from the measurement results. Given that an agent with a projective simulator lacks this additional information it does comparably well, and, conceivably, it can be improved by changing the update rule to incorporate more knowledge about the underlying quantum physics. For example, a positive measurement result and reward in one direction can be combined with a negative reward into the opposite direction, or, for each measurement result the reward is distributed according to how close all potential actions are to the rewarded one.

%----------------------------------------------------------------------
\subsection*{Adapting to Changing Fields and Improving Resolution}
\label{sec:adaption_to_changing_fields}
%----------------------------------------------------------------------

An important feature of the projective simulator is its ability to forget and thus to adapt to a changed situation. This ability distinguishes the present setting from schemes of parameter estimation, where the unknown parameter is assumed to be constant. For example, for a changing parameter standard Bayesian updating cannot be applied because past information needs to be disregarded and only recent information should be considered for estimating the current parameter. The projective simulator, in contrast, keeps track of an integrated average of past rewards for each action and is endowed with an element, the damping quantified by $\gamma$, to forget these rewards. The agent has the ability to completely change its behavior regardless of what has been rewarded earlier and irrespective of its earlier state. We shall consider two of such relearning scenarios in the following. We analyze the relearning by means of two quantities, the asymptotic success probability and the learning time it takes the agent to adapt.

%++++++++++++++++++++++++++++++++++++++++++++++++++++++++++++++++++++++
\begin{figure}[tbp]
	\includegraphics[width=0.98\linewidth]{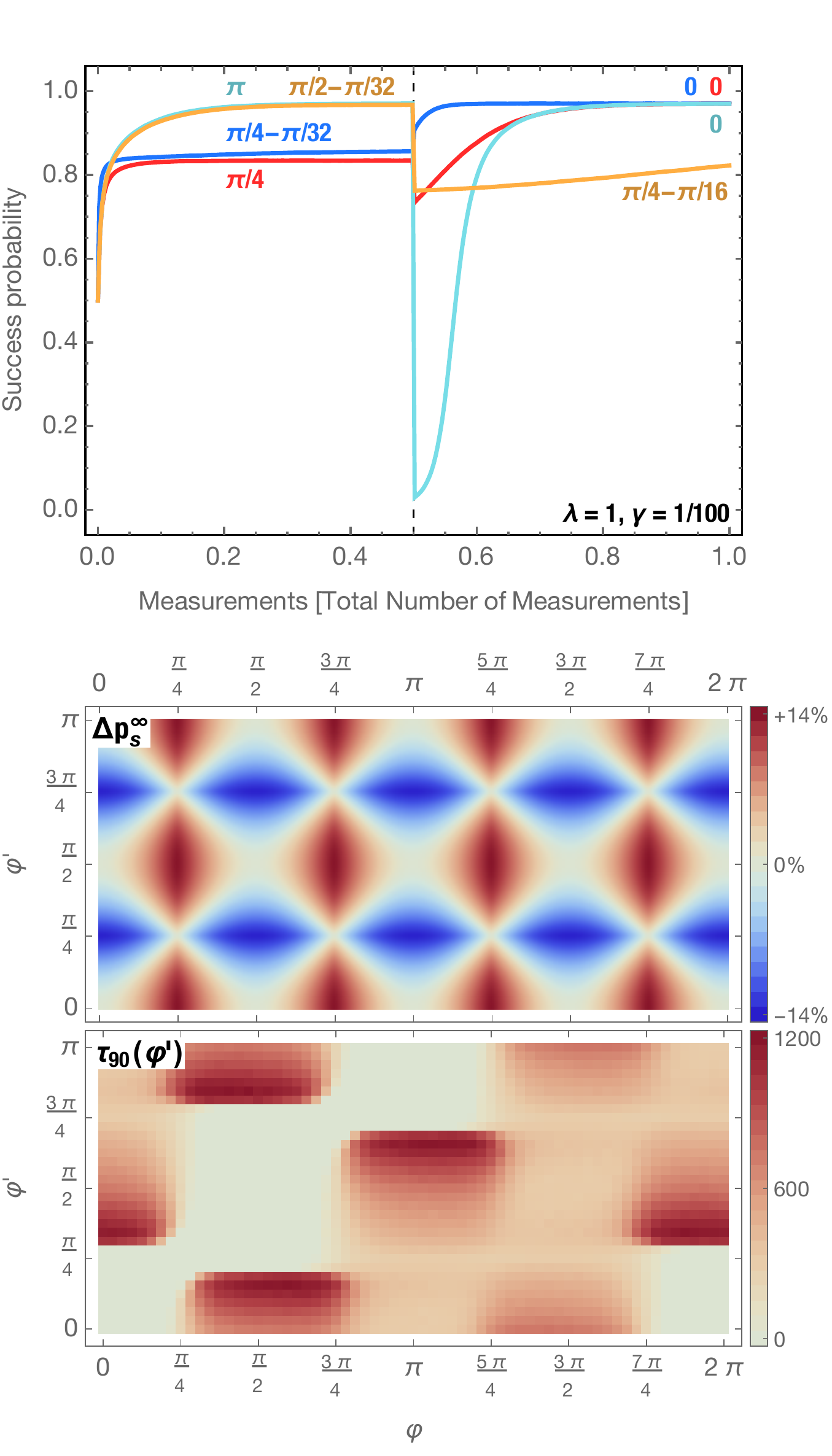}
	\caption{Relearning after a sudden change of $B$ leading to a shift from $\varphi$ to $\varphi^\prime$. An ensemble of 1000 agents first learns with $\varphi$ for $M/2$ measurements, after which the angle changes to $\varphi^\prime$ for another $M/2$ measurements, $\lambda=1$, $\gamma=1/100$.
	\textbf{Top:} Examples of learning curves for 4 different switches. $\varphi$ and $\varphi^\prime$ are given in the same color as the corresponding learning curve
	($M=10000$ for blue and $M=3000$ for the other examples).
	\textbf{Middle:} The asymptotic success probability (analytically obtained) shifts due to the change in angles.
	\textbf{Bottom:} The relearning time $\tau_{90}$ to reach 90\,\% of the asymptotic success probability after the field has changed also gives rise to a periodically repeating pattern and shows a structure commensurate with the choice of projectors in intervals of $\pi/2$.}
	\label{fig:Relearning}
\end{figure}
%++++++++++++++++++++++++++++++++++++++++++++++++++++++++++++++++++++++

\textit{Relearning After a Switched Field.}---
Changes of $B$ result in a different $\varphi$ and require the agent to adapt and relearn. For a single sudden change in $B$ the angle $\varphi$ changes only once at a certain time to a new angle $\varphi^\prime$. Depending on the values of $\varphi$ and $\varphi^\prime$ the agents shows a rich landscape of relearning patterns as illustrated in Figure~\ref{fig:Relearning}.

After the switch the asymptotic success probability is always that of the new $\varphi^\prime$ and may lie above or below the success probability of the old $\varphi$ (Figure~\ref{fig:Relearning} top). The change in $p_s^\infty$ is illustrated in Figure~\ref{fig:Relearning} (middle). The sudden drop or increase in success right after the change of the angle and the time to reach a success probability depends strongly on the relation of the two angles and how much of the internal state ($h$-values) needs to be changed to reach the new state. These effects in the relearning time appear in addition to the known effects of changing the reward scaling $\lambda$ and damping rate $\gamma$~\cite{PS,PScompreh}. A summary of the relearning times and change in asymptotic efficiencies is given in Figure~\ref{fig:Relearning} (middle and bottom).

\textit{Time-dependent Fields.}---%
An important feature for applications is the agent's ability to adapt its actions to slowly changing external fields. An agent's state is the result of a dynamical equilibrium between rewarded actions in the past and forgetting this information on a time scale given by $\gamma$. Therefore, the speed at which an agent can adapt is limited by the speed with which it can modify its internal state. The agent can adapt to a change in the reward landscape caused by a changing field as long as it has enough time to sample the modified reward landscape and modify its internal state accordingly, which depends on $\lambda$ and the timescale given by~$\gamma$.

\begin{figure}[tbp]
	\includegraphics[width=\linewidth]{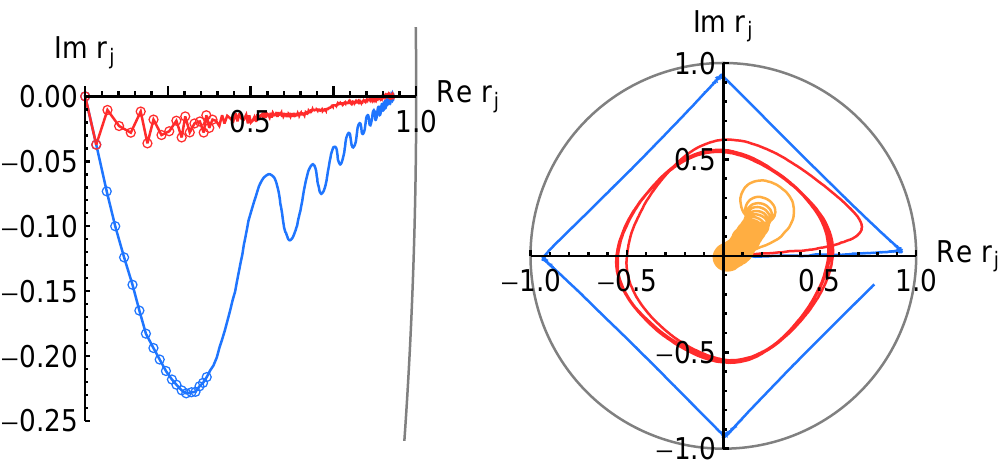}
	\caption{Adaptation to time-dependent magnetic fields: Trajectories of the state vector $\bar{\vec{r}}=\sum_\alpha \bar{p}_\alpha \e^{i \alpha}$ as averaged over an ensemble of 1000 agents. The state vector of the ensemble starts in the origin.
	\textbf{Left:} For fast oscillating fields $\varphi(n)=-\frac{\pi}{4}\cos(\omega n)$ with $\omega=10$ (red) and $\omega=1/10$ (blue) the agents adapt to the average angle $\varphi=0$ over 5000 measurements. Data points are explicitly indicated for the first 20 measurements and joined by a line.
	\textbf{Right:} Linearly drifting fields $\varphi(n)=\omega n$ can be learned by the agent the better the slower they change on the timescale of the learning time: $\omega=\pi/5000$ (blue) shown for 10000 measurements, and $\omega=\pi/500$ (red), $\omega=\pi/10$ (orange) shown for 4000 measurements each. The ensemble follows the field and the state trajectory converges to a limiting cycle.}
	\label{fig:TimeDependentFields}
\end{figure}

Figure~\ref{fig:TimeDependentFields} shows two examples.
The first example (left) is a setting with a fast oscillating field, i.e., one with $\varphi(n)\sim \cos(\omega n)$ as a function of the measurement round~$n$, where only the time average is learned because the agent effectively takes samples from the entire reward landscape. The state vector converges to angle $\varphi=0$ and reaches almost unit length.
The second example (right) shows a setting with a linearly increasing magnetic field, giving rise to $\varphi(n)\sim n$. As $\varphi$ moves anticlockwise on the unit circle as a function of the measurement round, the agent can keep up as quantified by $\bar{\vec{r}}$ with a state trajectory that also moves counterclockwise albeit with a slight delay and the length of the state vector is longer, i.e., the field is learned better, for a slower rate of change.

%++++++++++++++++++++++++++++++++++++++++++++++++++++++++++++++++++++++
%++++++++++++++++++++++++++++++++++++++++++++++++++++++++++++++++++++++

%----------------------------------------------------------------------
\subsection*{Initial Choice of Measurement Directions and Composition}
\label{sec:composition_techniques}
%----------------------------------------------------------------------

The choice of projectors that the agent can measure affects the agent's success in two ways: On one hand it fixes the available angles and thereby ability to measure the correct angle. A finer grained sample of measurement angles is beneficial because it will contain an angle that is closer to the actual angle and allow for almost perfect measurements. It also avoids efficiency minima due to the coarse-graining as they appear in Figure~\ref{fig:learning_single_phi} (top, right). A finer resolution of measurement angles, however, will introduce many angles that are almost equally successful and are hard to distinguish by their average reward.
On the other hand, the choice of measurement angles fixes the discrete support on which the probability distribution for $\varphi$ can be built, which contains the information on the angle $\varphi$. A drawback of a coarse-grained support is the arising large variance in the distribution. A fine-grained support, however, needs lots of sampling to evaluate each individual point in the distribution.

%++++++++++++++++++++++++++++++++++++++++++++++++++++++++++++++++++++++
\begin{figure}[tbp]
	\includegraphics[width=\linewidth]{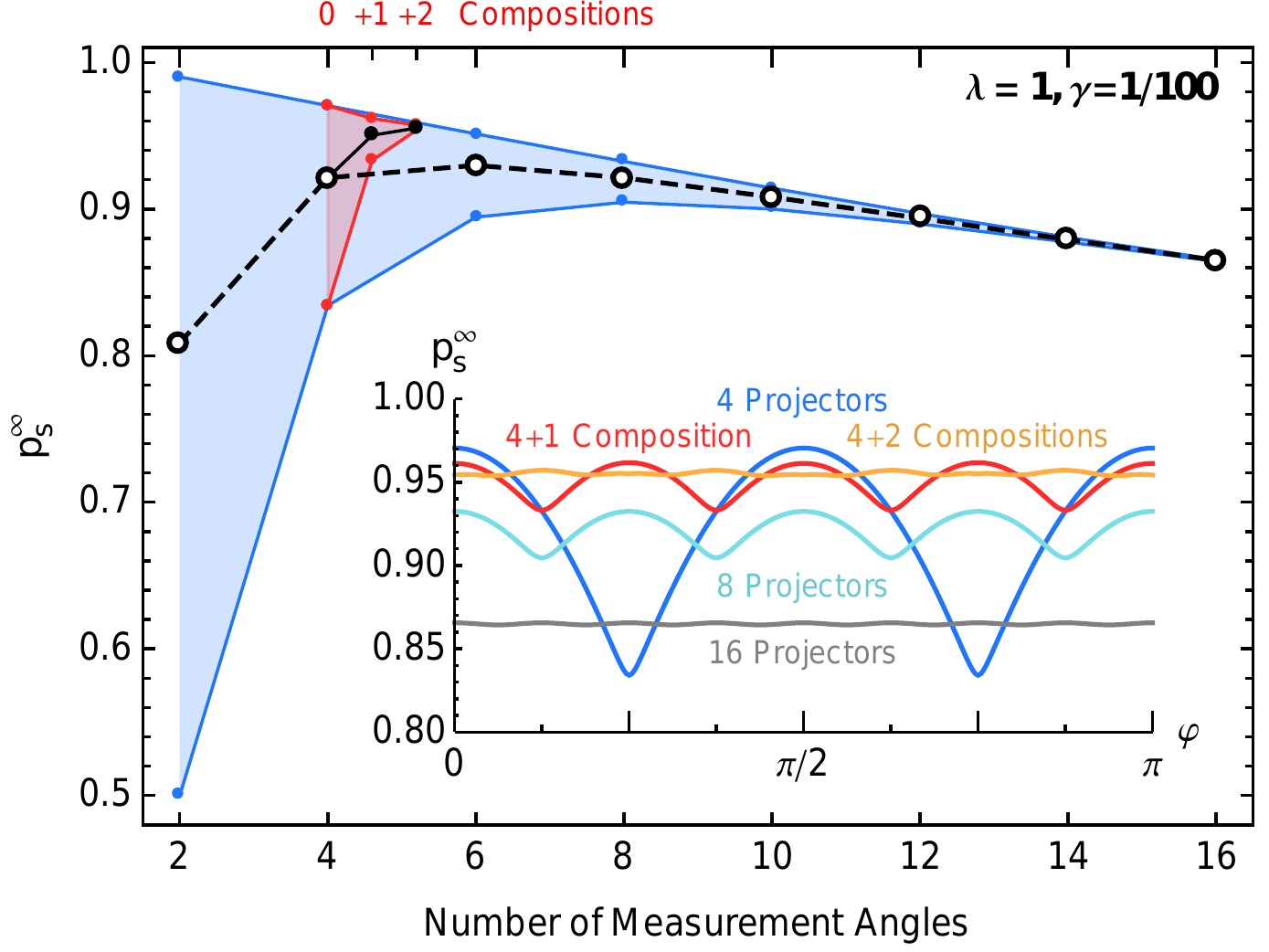}
	\caption{Success probability for $2N_\alpha$ measurement directions initially available to an agent (blue), uniformly spaced on the circle, and with additional bisection compositions (red). Data points are angular averages, and the vertical region denotes the maximum and minimum of the success probability. These data summarize the angular dependence of $p_s^\infty$ as depicted for several examples in the inset.
	The reason for a decrease of $p_s^\infty$ with an increase in measurement directions is due to damping.}
	\label{fig:InitialProjectorsComposition}
\end{figure}
%++++++++++++++++++++++++++++++++++++++++++++++++++++++++++++++++++++++

As there are advantages and disadvantages to the number of measurement directions, we ask if there is an optimal number of fixed projectors. In order to distinguish directions on a circle, at least three directions are needed. (For just two directions that are equally successful, it is not possible to decide which of the two angles between those two directions is the correct one.)
We have calculated the asymptotic success probability for an agent that has access to $2N_\alpha$ measurement angles that are equidistantly spaced on the unit circle, i.e., it can measure $N_\alpha$ observables of the form \eqref{eq:observable} with two eigenstates each that are opposite on the equator of the Bloch sphere. The angular dependence of the success probability in Figure~\ref{fig:InitialProjectorsComposition} (inset) shows maxima at angles that are measurement directions and minima between two neighboring angles. As more observables are added, the worst cases in between two neighboring projectors improve, but at the same time also the optimal cases decrease because the optimal angle is not chosen as frequently due to slightly better neighboring angles that are also rewarded more often. The success probability averaged over all angles first increases and then decreases as summarized in Figure~\ref{fig:InitialProjectorsComposition}. In the limit of large $N_\alpha$ the success probability converges to 50\,\% because of the constant damping $\gamma$.
In the example case with $\lambda=1$ and $\gamma=1/100$ we can give these recommendations: For optimizing the best case, the number of 2 projectors is optimal, for the best worst case success probability, 8 projectors are best, and for the best average success probability 6 projectors are the best initial choice.

A strategy to mitigate the decrease in overall efficiency for a more refined angular resolution is \emph{composition}, which is one of the original features of projective simulation~\cite{PS,PScompreh}. With composition the projective simulator is endowed with the ability to generate new clips based on the composition of already existing ones. For parameter estimation the projective simulator can insert new clips with new measurement directions only where additional resolution is needed.

The composition mechanism is an additional dynamical element in the projective simulator. Based on the state of the projective simulator it is triggered and inserts a new clip based on existing ones. These new elements, i.e., the trigger mechanism, constructing the new clip, and how the new clip is inserted into the network must be specified and leave room for arbitrarily complicated rules. We will restrict to the simplest mechanisms, which will also draw some intuition from actual conceivable physical dynamics.

%++++++++++++++++++++++++++++++++++++++++++++++++++++++++++++++++++++++
\textit{Bisecting Composition.}---%
The first composition mechanism simply operates by bisection and refining the resolution in the relevant regions. After the agent has learned with its initial set of projectors, the two actions clips with the largest $h$-values are selected and used to compose a new clip between the two. In situations with angles $\varphi=\pi/4$ or $\pi/8$ the action clips with $\alpha=0$ and $\pi/2$ will have the largest $h$-values and give rise to the creation of a new clip with $\alpha=\pi/4$, which improves the resolution of the discretization in the first quadrant.
The success probability before and after one such composition is depicted in the inset in Figure~\ref{fig:InitialProjectorsComposition} as light blue and red curve, respectively. For the angle $\varphi=\pi/4$ the success probability is increased from a minimum of 83.4\,\% to a maximum with 96.2\,\% without adding unnecessary projectors in the remaining quadrants, which would lower $p_s^\infty$ to 93.2\,\% of the curve with 8 projectors.
When always adding a single additional angle in the middle of the quadrant in which $\varphi$ lies, worst case scenarios for $p_s^\infty$ appear only for angles like $\pi/8$ and $3\pi/8$ with 93.32\,\%, which still is a slight improvement over the coarse graining with only 4 projectors (93.26\,\%). For $\varphi=\pi/8$ the composition at $\alpha=\pi/4$ is helpful but suboptimal. For angles at the projectors, e.g. $\varphi=0$, an additional composition is harmful and decreases $p_s^\infty$ from 97.1\,\% to 96.1\,\%.
A single composition that doubles the angular resolution in one quadrant is qualitatively similar to 8 initial measurement angles, but with a higher success probability.
A second composition step that adds another projector with an angle of odd multiples of $\pi/8$, effectively reproduces the resolution of 16 initial angles but only in one octant of the unit circle. It improves the worst cases at the cost of a slightly reduced overall success probability. Even more bisections will further increase the angular resolution but reduce the overall success to the point that they are counterproductive.
Although a bisecting composition is very simple approach, it provides the advantage of a larger number in initial projectors while avoiding a large penalty in overall efficiency due to a large action space with the same parameters.

%++++++++++++++++++++++++++++++++++++++++++++++++++++++++++++++++++++++
\textit{Composition with the Glow Mechanism.}---%
The second mechanism departs from the strict bisection strategy of the first mechanism. The agent reaches an optimal success probability if it can measure along the direction $\alpha=\varphi$. The bisection strategy only approximates $\varphi$ and sometimes introduces unnecessarily many angles, e.g., for $\varphi=\pi/8$ the additional angle $\alpha=\pi/4$ has to be built first. We overcome this disadvantage by a better use of the information provided by the measurement results to estimate which new projector angle should be inserted as addition action.
We employ a variant of the ``edge glow mechanism''~\cite{PScompreh} to compose a single new action clip in the following way. We assign a second degree of freedom to each edge called ``glow'' and denote it by $g_\alpha$. Instead of updating the $h$-values with the reward according to \eqref{eq:updateRule}, we first accumulate rewards in the $g_\alpha$ according to the following update rule:
\begin{align}
	h_{\alpha^\prime}(n+1) &= h_{\alpha^\prime}(n), \\
	g_{\alpha^\prime}(n+1) &= g_{\alpha^\prime}(n) +  \delta_{\alpha \alpha^\prime} \lambda_n,
\end{align}
with initial values $g_\alpha(0)=0$. The change in the update rule for $h$ effectively amounts to setting $\lambda=0$ and $\gamma=1$.
The behavior of the agent remains unchanged as the $h$-values remain at their initial values $h_\alpha(0)=1$, i.e., the agent measures equally often in all available directions. However, since there is no bias in the frequency of available measurement direction, the accumulated rewards in the respective $g_\alpha$ provide a measure of the average reward for each direction. Once the agent sampled enough measurement results, e.g., when the first $g_\alpha$ surpasses the threshold $g_\text{thresh}=500$, a new action clip is composed and inserted into the projective simulator. The new measurement direction $\bar{\alpha}$ is composed from all $\alpha$ and weighted by the $g_\alpha$:
\begin{equation}
	\bar{\alpha} = \arg \sum_\alpha g_\alpha \e^{i\alpha},
\end{equation}
and we set the new $h_{\bar{\alpha}}=\sum_\alpha g_\alpha$.

In order to prevent that a direction is inserted that is already present, the agent first checks that $\bar{\alpha}$ is sufficiently different from all already existing $\alpha$, e.g., by inserting $\bar{\alpha}$ only if it differs from $\alpha$ by more than $1/10$ circular standard deviations of the circular distribution given by the $g_\alpha$. If $\bar{\alpha}$ is too close to one $\alpha$, the $h_\alpha$ of this $\alpha$ is instead strengthened and set equal to the sum of all $g_\alpha$, and no new clip is inserted.
After the composition, we continue with the usual update rule for the $h$-values.

%++++++++++++++++++++++++++++++++++++++++++++++++++++++++++++++++++++++
\begin{figure}[tbp]
	\includegraphics[width=\linewidth]{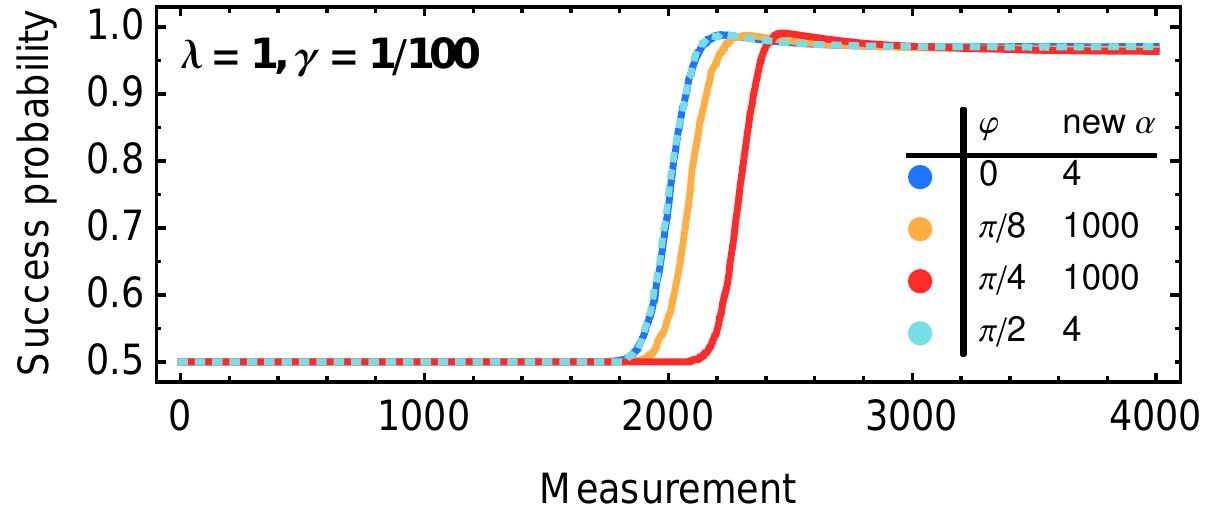}
	\caption{Learning curves for glow composition averaged over an ensemble of 1000 agents for various $\varphi$. The $h$-values are only updated with $\lambda$ and $\gamma$ after the composition. For $\varphi$ coinciding with an existing $\alpha$ only 4 agents compose a new angle $\bar{\alpha}$, which is more than $\sigma/10$ away from an existing $\alpha$, whereas all agents compose angles for the other examples of $\varphi$. The position of the step depends on the choice of the threshold for composition, here $g_{\text{thresh}}=500$, which is chosen for large statistics but can be decreased without much penalty in the asymptotic efficiency albeit at the cost of slightly less accurate composed angles.}
	\label{fig:glow}
\end{figure}
%++++++++++++++++++++++++++++++++++++++++++++++++++++++++++++++++++++++

The learning curves for the this form of glow composition are shown in Figure~\ref{fig:glow}. Starting with 4 angles and a $g_\text{thresh}=500$, at least 2000 measurements have to be done on average before the first composition can occur. This threshold can be decreased leading to a faster composition, albeit at worse statistics, which result in inaccuracies of the composed angles. Inaccurate compositions, however, impact the success probability only to a small extent because it decreases with the cosine of the angular difference between $\varphi$ and the composed angle.

In the direction of an existing angle, e.g., $\varphi=0$ or $\pi/2$, the first amplifications of the respective $\alpha$ occurs starting with 2000 measurements. For the direction $\varphi=\pi/4$, more measurement need to be done on average to reach $g_0$ or $g_{\pi/2}=500$ because these direction are not rewarded with certainty, and composition occurs on average later, with $\varphi=\pi/4$ being one of the four latest instances. After the composition the success probability jumps from 50\,\% to about 99\,\%.
Since the newly set $h$-value for the best measurement direction is larger than the steady-state value for our choice of $\lambda=1$ and $\gamma=1/100$, the success probability decreases slightly to approach $p_s^\infty$ from above.

In an ensemble of 1000 agents only 4 compose an angle when $\varphi=0$ or $\pi/2$, whereas all do a composition for $\varphi=\pi/8$ or $\pi/4$. In our numerical experiment, the distribution of composed angles $\bar{\alpha}$ is sharply peaked around $\varphi$ with a $\sigma\approx\pi/100$.

By using the glow mechanism to obtain an effective average reward for each measurement direction, and then composing a mean angle from the reward distribution, the agent effectively creates a weighted sum of directions. It thereby embodies a method similar to the estimation of expectation values done in state tomography.

%----------------------------------------------------------------------
\subsection*{Adapting multiple measurement directions}
\label{sec:multiple_directions}
%----------------------------------------------------------------------

So far we have demonstrated how an agent equipped with a suitable projective simulator can align a single measurement direction, e.g., $\hat{\sigma}_x$ for the state $\ket{+}$, with an initially unknown state $\ket{\varphi}$, which emerged from $\ket{+}$ due to a magnetic field. Since one of the aims is to employ the agent as a means to carry out measurement-based quantum computation (MBQC)~\cite{MBQC} in an unknown external field, all measurement directions that are required to run a specific algorithm in MBQC need to be adapted to this unknown stray field. We therefore need to extend the projective simulator to learn several measurement directions, which shall be given as the respective input. Ultimately, the agent would translate the measurement directions necessary for the algorithm to the reference frame that rotates due to the magnetic field.

We modify the inital agent setup depicted in Figure~\ref{fig:agentSetup} in the following way. The step that prepares the defined initial state $\ket{+}$ is removed and the qubit is simply left in the state that is prepared by the previous measurement. The projective simulator now receives as an input not just a trigger event, which activated the $*$-clip, but now it receives the previous measurement direction and the obtained measurement result as a percept. The initial state and percept can be chosen arbitrarily, e.g., at random, as they do not matter in the subsequent feedback loop.

%++++++++++++++++++++++++++++++++++++++++++++++++++++++++++++++++++++++
\begin{figure}[tbp]
	\includegraphics{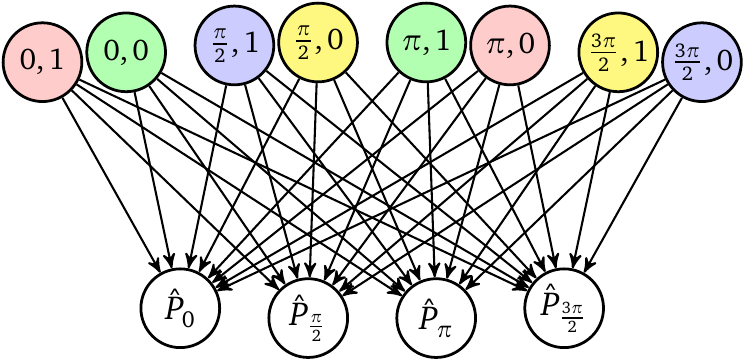}
	\caption{Clip network of the extended projective simulator. Percepts are all possible combinations of the previous measurement direction $\alpha$ and obtainable measurement outcome (0 or 1). Percepts that correspond to the same state prepared by the previous measurement are colored equally.}
	\label{fig:network_8to4}
\end{figure}
%++++++++++++++++++++++++++++++++++++++++++++++++++++++++++++++++++++++

For the new scheme, we also extend the clip network of the projective simulator to 8 percept clips, which represent all combinations of previous measurement direction and obtained reward, as depicted in Figure~\ref{fig:network_8to4}. Effectively, the extended clip network consists of 8 copies of the previous simple clip network, which are activated according to the actions and results of the previous time step. The agent enters a feedback cycle, where measured directions and outcomes are fed back to the agent. The information about which state preparation method was used is available to the agent as percept, and thereby it indirectly receives a hint about which state has been prepared. Given each prepared state, which then evolves to acquire an additional shift in the angle by $\varphi$, the agent learns which measurement direction most likely matches this rotated initial state.
To give an example, consider the test qubit in the initial state $\ket{+}\equiv\ket{\alpha=0}$, which evolves into $\ket{\varphi}$. The agent measures this state, say along $\alpha=\pi/2$, and obtains result $1$. It thereby prepares the test qubit in state $\ket{\pi/2}$, which again evolves for time $\Delta t$ into $\ket{\pi/2+\varphi}$ for the next measurement. This next measurement is chosen according to the $h$-values of edges originating from the percept clip ``$\pi/2,1$'' to each of the four actions.

%++++++++++++++++++++++++++++++++++++++++++++++++++++++++++++++++++++++
\begin{figure}[tbp]
	\includegraphics[width=\linewidth]{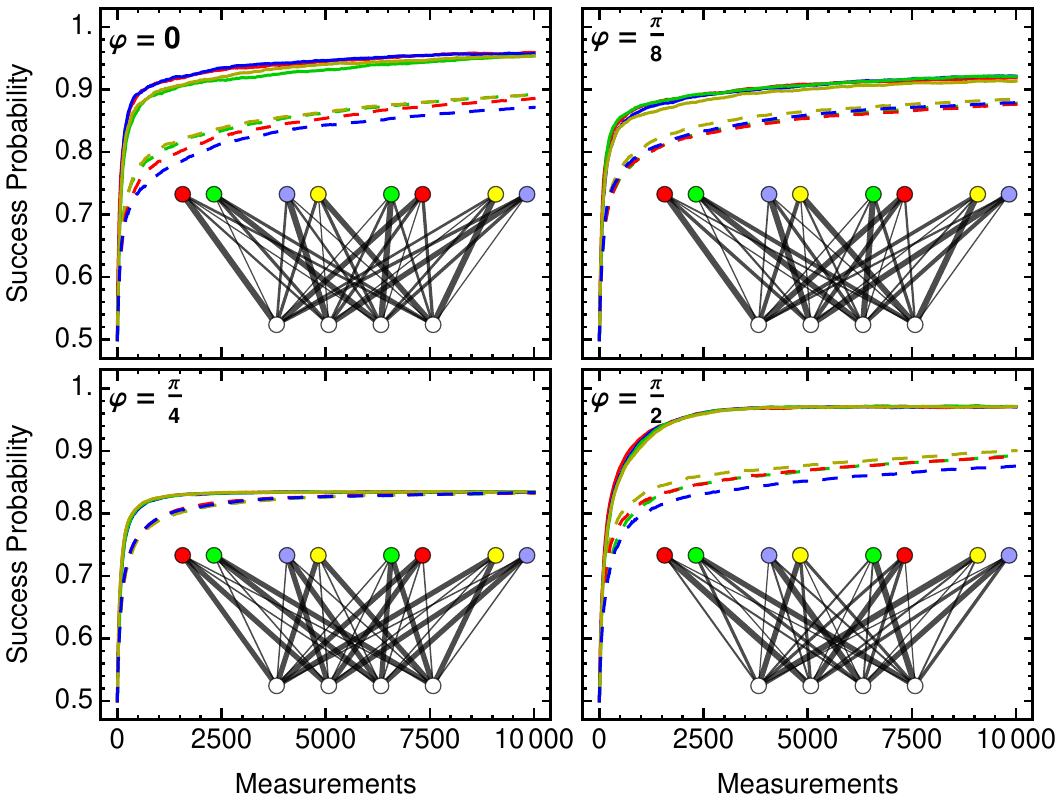}
	\caption{Learning curves of a projective simulator with 8 percepts (state preparation procedures) and 4 actions (measurement directions) for various $\varphi$. Plotted are conditioned success probability, i.e., given a percept, what is the probability of obtaining a +1 measurement outcome, where each curve corresponds to one percept, solid lines represent outcome-$1$ preparation methods, dashed lines those for outcome~$0$. Color codings are the same as for the clip network. Curves are averages over 100 agents, with $\lambda=1$ and $\gamma=1/100$. For the last time step the clip network with $h$-values encoded in the thickness of the edges are given in the inset. Colors and clips match those of Figure~\ref{fig:network_8to4}.}
	\label{fig:completePSLearning}
\end{figure}
%++++++++++++++++++++++++++++++++++++++++++++++++++++++++++++++++++++++

The clip network is now much larger than before and the agent needs more measurements to update all the connections until the $h$-values converge into those of the steady state. Naively, we can expect an 8-fold increase, however, since the agent converges to a state in which measurements that give outcomes $1$ are preferred, learning the right measurements for a outcome-$0$ preparation is delayed.
This learning behavior is shown in Figure~\ref{fig:completePSLearning}, where outcome-$1$ preparations converge early and outcome-$0$ preparations later, which in turn also delays the overall convergence. Naturally, the training of the whole network is faster in situations where the $0$ outcomes occur more often, e.g., for $\varphi=\pi/4$, or in situations that lead to different measurement directions, e.g., for $\varphi=\pi/2$.

As the agent encounters situations with different percepts, the number of time steps in between two successive activations of the same percept is now increased on average. This leads to a qualitative and quantitative change in the learning curves as compared to the previous simple agent with only one percept. The number of times that the damping reduces the $h$-value of each edge would increase and lead to a reduced efficiency, because the agent forgets too quickly in between rewards. To maintain high $h$-values for rewarded transitions we could adjust $\gamma$ to a lower value, but we choose to simply restrict the application of the update rule, and the application of the damping in particular, to a subgraph of the clip network, namely, only those edges that are connected to the activated percept clip. Thereby we maintain the quantitative behavior of the simple clip network used in the previous sections.

Percepts give the preparation procedure of the test qubit and thereby effectively encode information about which state has been prepared. A closer inspection reveals that each state is represented twice because is can be prepared in two ways, e.g., $\ket{+}$ can be prepared by a measurement of $\hat{P}_0$ with outcome 1 or by $\hat{P}_\pi$ with outcome 0. Preparation procedures that result in the same prepared states are highlighted with the same color of the percept clip in Figures~\ref{fig:network_8to4} and \ref{fig:completePSLearning}. This redundancy increases the learning times because the same behavior has to be learned twice. The clip network could be optimized with an additional intermediate layer that first maps preparation methods to states, which may be learned first without a stray field, and then the prepared states to best measurement directions in a stray field.

Once the agent has adapted its measurement directions to the unknown external field with a test qubit, it can be used as a translator between intended measurement directions and their corresponding directions in the rotated reference frame. This application of a trained agent works as follows. After a training period, we  fix all the $h$-values. Instead on the test qubit, the agent now acts on the qubit that needs to be measured along a certain direction according to a MBQC scheme, for example. We then excite a percept of the agent that corresponds to the direction of the intended measurement direction in zero field. The agent then chooses most likely the measurement direction that corresponds to this measurement in the rotated frame, i.e., the measurement that takes the field into account.

%----------------------------------------------------------------------
\subsection*{Measurement-based Grover Algorithm}
\label{sec:measurement_based_grover_algorithm}
%----------------------------------------------------------------------

We first briefly repeat the MBQC variant of the Grover search algorithm for a database with 4 elements~\cite{GroverPhotonsMBQC} and adapt it to our notation and use of projective measurements $\hat{P}_\alpha$.
The initial resource state is a cluster state of 4 qubits in ring form, i.e., starting from the state $\ket{+}^{\otimes 4}$ we apply a controlled phase gate between the qubit pairs 1--2, 2--3, 3--4, 4--1, and obtain
\begin{align}
	\ket{\Psi_0} = \frac{1}{2} \Big(
		& \ket{0}\ket{+}\ket{0}\ket{+}
		+ \ket{0}\ket{-}\ket{1}\ket{-} \nonumber \\
		+& \ket{1}\ket{-}\ket{0}\ket{-}
		+  \ket{1}\ket{+}\ket{1}\ket{+}
		\Big).
\end{align}
A database with 4 entries (i.e., with elements $00$, $01$, $10$, and $11$) only requires a single Grover step to find the marked element.
The algorithm starts by doing this one necessary query to the database and thereby marks the database entry that is to be found. A measurement of the projectors $\hat{P}_0$ or $\hat{P}_\pi$ on qubits 1 and 4 realizes the specific database, where each pair of measurement directions $00$, $0\pi$, $\pi0$, and $\pi\pi$ corresponds to marking the database element $00$, $01$, $10$, and $11$, respectively. For each of the two measurements of $\hat{P}_0$ or $\hat{P}_\pi$ both measurement results $r_{1,4}=0$ or $1$ appear with probability $1/2$. Therefore, the results alone do not allow us to infer the measurement directions and thereby the marked element. In the problem setting of the algorithm the choice of measurement directions is hidden. Only from the measurement results of qubits 1 and 4, and from the measurements done on the remaining two qubits, we should infer the marked element. On the remaining qubits we therefore measure the observable $\hat{P}_0$, whose measurement outcome depends on the measurement directions on qubits 1 and 4, and is correlated to the previous two outcomes. Finally, the calculation of $(r_1\oplus r_3, r_2\oplus r_4)$, i.e., addition of the measurement outcomes modulo 2, reveals the two bits of the marked element with certainty.
Although, at the present point the MBQC version of Grover's algorithm appears to merely uncover (anti-)correlations between measurement directions, there is an explicit mapping between the quantum circuit of Gover's algorithm on one hand, and the circuit for creating and measuring the cluster state on the other~\cite{GroverPhotonsMBQC}.

%++++++++++++++++++++++++++++++++++++++++++++++++++++++++++++++++++++++
\begin{figure}
	\includegraphics[width=\linewidth]{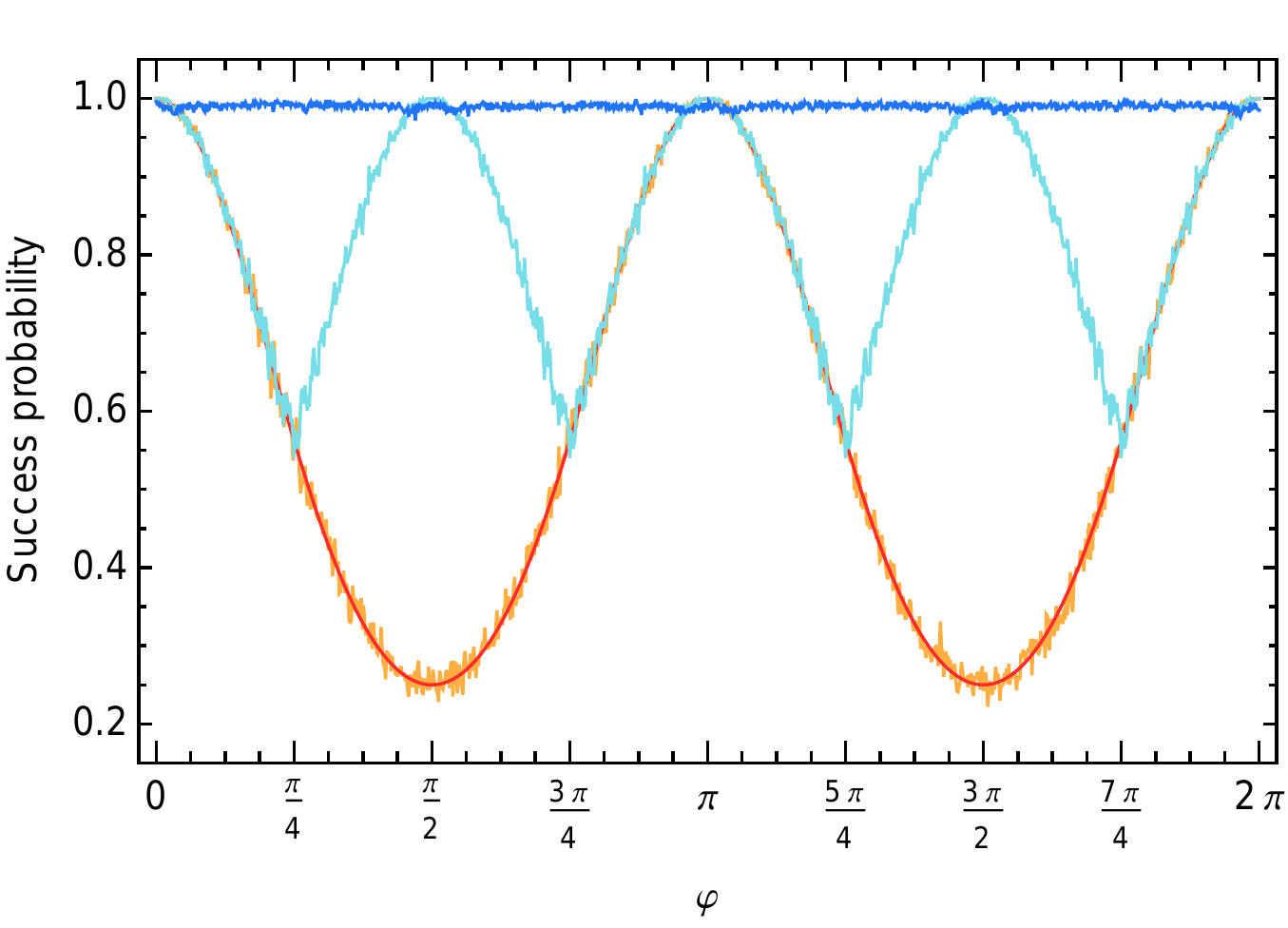}
	\caption{Fidelity of the Grover search algorithm in the presence of an unknown static magnetic field that rotates every qubit by angle $\varphi$ along the equator of the Bloch sphere. Data is obtained in independent runs with marked element $00$ for all fields giving rise to angles $\varphi$ between 0 and $2\pi$ in steps of $\pi/500$. Noisy data is the fraction of an ensemble of agents that identifies the marked element correctly when performing all four measurements in Grover's algorithm.
	Red (analytical) and orange (numerical, 3000 agents) curves give the success of the Grover search (all four measurements) without taking into account the field in the measurement direction.
	The light blue curve gives the success for an ensemble of 1000 agents that each have a perfectly trained projective simulator with 4 measurement directions, which has learned the external magnetic field before doing the measurements for the Grover search algorithm.
	The dark blue curve is an ensemble of 1000 agents that employs the glow mechanism to build a measurement direction that is adapted to the external magnetic field before using it to perform the Grover search.
	}
	\label{fig:Grover}
\end{figure}
%++++++++++++++++++++++++++++++++++++++++++++++++++++++++++++++++++++++

If the initial state is placed in an unknown external field pointing along the $z$-direction, the state $\ket{\Psi_0}$ is transformed into $U\otimes U\otimes U\otimes U\ket{\Psi_0}$ with the local unitary rotations $U=\exp(-i \varphi \hat{\sigma}_z / 2)$. If we recall that $U\ket{+}=\ket{\varphi}$ it is straightforward to see that the measurement protocol of the Grover algorithm will no longer give the correct marked element because the external field effectively shifts the measurement directions by the angle $-\varphi$ with respect to the original measurement directions. As a result the probability to identify the correct marked element, i.e., the success probability of the algorithm, is periodically modulated by $\varphi$ and a straightforward calculation gives
\begin{equation}
	p_s(\varphi) = \frac{1}{16} \big(3+\cos 2\varphi\big)^2.
\end{equation}
For $\varphi$ being a multiple of $\pi$ the algorithm works perfectly because the local rotations align the qubits' reference frames again with the $x$-axis. The Grover algorithm is invariant under the inversion of all measurement directions (i.e., changing all directions $0$ to $\pi$ and vice versa).
In the worst case, for $\varphi$ being an odd multiple of $\pi/2$, the chance of identifying the right element is 1/4, as the measurements of the intended Grover search actually reveal no useful information because they are unbiased with respect to the required measurement direction. In Figure~\ref{fig:Grover}, the analytic results for identifying the correct marked element $00$ in the rotated state match the trials with 1000 agents that simply measure all 4 qubits the direction $\alpha=0$ and then try to identify the marked element from the obtained measurement results.

For testing the agent with a projective simulator we restrict to a realization with a single marked element, namely $00$, which can be implemented with measurements along the $x$-axis, all in the direction $\alpha=0$.
The agent first learns with a test qubit exposed to the external field and adapts to the field strength. We then fix these obtained $h$-values and use the agent without update rule to carry out the four measurements on the cluster state, one after the other, according to its available measurement directions and internal probabilities.

The first example is an agent that has only 4 fixed measurement directions available ($\alpha$ being a multiple of $\pi/2$), which we first train to achieve optimal success probability with the test qubit. The optimal performance is reached in the limit $\gamma\to0$, which amounts to $p_\alpha=1$ for the single $\alpha$ that is closest to $\varphi$ and all others zero. The light blue curve in Figure~\ref{fig:Grover} shows the fraction of the agent ensemble that identifies the marked element with this projective simulator correctly. The Grover search is recovered perfectly for fields with $\varphi$ being a multiple of $\pi/2$, which can be matched exactly by the available measurement directions.

The second example is an agent that first learns with a test qubit in the external field with composition according to the glow mechanism. That is, after 2000 measurements on average, the agent composes a new measurement direction or strengthens an existing one that matches $\varphi$. The $h$-values after the composition remain fixed and the agent measures the cluster state according to the available measurement directions and probabilities. The dark blue curve in Figure~\ref{fig:Grover} illustrates that an ensemble of this kind of agent is highly successful in doing the Grover search for all angles $\varphi$. The shortfall from a perfect performance (the average success probability is 99.0\,\% with a standard deviation of 0.3\,\%) originates in the slight deviations of the composed angle from $\varphi$ and the non-zero probability to chose the remaining non-optimal measurement directions.

%======================================================================
\section*{Discussion}
\label{sec:Discussion}
%======================================================================

We presented an autonomous adaptive system that is able to perform quantum information processing in changing environments. The controller is a learning agent endowed with a projective simulator that adapts measurement directions in a setup of measurement-based quantum computation by reinforcement learning. Our approach thus combines elements from embodied artificial intelligence with the purpose of carrying out robust quantum information processing.

In an exemplary setup of adapting measurement directions to an unknown stray magnetic field in a fixed direction, we have characterized the learning process of the projective simulator and its adaption to time-varying fields using numerical studies. We found that an agent using projective simulation is able to adapt to such unknown stray fields.
We provide analytical estimates of its success probability in limiting cases of the non-linear learning process.
In our scenario the agent may adapt the measurement direction by drawing from a initially provided set of fixed measurement directions. We have characterized the performance of the agent for different sets of available measurement directions and we explored composition mechanisms to create new and better measurement directions on the fly, together with the corresponding internal structure in the projective simulator. Strategies with composed measurement directions surpass strategies with fixed sets of directions in both learning speed and resulting efficiency.
As a demonstration of adaptive quantum information processing, the agent successfully carries out a measurement-based version of Grover's search algorithm in the presence of a detrimental unknown external magnetic field.

The present approach can be readily extended and improved in several directions as indicated in the respective sections in the paper.
First and foremost, the agent effectively develops and embodies rules to cope and operate with quantum mechanical systems, which are seeded by the specific form the update rule together with the reward scheme, and the composition mechanisms. Both of these elements start from simple primitives, e.g.\ ``prefer a specific measurement if it more likely results in a +1 measurement outcome'' for the update rule, and give rise to a sensible and sufficient behavior in our problem setting. Both can be improved by effectively incorporating more information about the quantum mechanical nature of the underlying problem domain, however, at the expense of more complicated update and composition rules.
Errors or imperfect measurements can be straightforwardly incorporated into the present scheme by using POVMs instead of projective measurement, or by adding a classical noise, e.g. bit flips, to the measurement outcomes. Such errors lead to a diluted information about which measurement directions are correct and give $+1$ measurement outcomes. In the presence of errors, spurious rewards appear for wrong measurement directions and the average reward for correct measurement directions is reduced. Both effectively diminish the contrast in the reward landscape, which is equivalent to a lower reward scaling factor $\lambda$. We expect that the agent will still be able to learn in such situations, but it will take longer to do so and reach a lower asymptotic success probability. The latter can partly be recovered by adjusting $\lambda$ and $\gamma$, however, an increase of the learning time over a noiseless scenario will remain.

The long-term goal of this investigation is to develop integrated and autonomous schemes for measurement-based quantum information processing that can adapt to changing environments. In our scheme, learning is not realized by feedback from some external macroscopic sensor, e.g. a magnetometer, but it uses only information drawn from measurements on qubits, which are also the operations that drive the processing of the quantum information. In this sense our approach is related to recent work on intelligent quantum error correction~\cite{estimationFromSyndromes}.

The approach that we have presented in the present paper can be generalized and integrated into a scheme of universal measurement-based quantum computation, where measurements of stabilizer operators of a cluster state are used both for the correction of errors on the resource state and, at the same time, for the adaption of measurement directions that drive the quantum computation. This will be reported elsewhere.

We note that the projective simulator does not assume that rewards originate from measurement probabilities of a quantum state and, therefore, it is ``model free''. This also opens the path to study foundational questions such as, to what extent can a machine effectively learn the rules of quantum mechanics through simple reinforcement processes.

%======================================================================
%% Acknowledgments
%======================================================================
\begin{acknowledgments}
We thank Wolfgang Dür for initial discussions on this topic and Vedran Dunjko for comments on the manuscript.
We acknowledge support from the Austrian Science Fund (FWF) through the SFB FoQuS: F4012, and the Templeton World Charity Foundation grant TWCF0078/AB46.
\end{acknowledgments}

%%%%%%%%%%%%%%%%%%%%%%%%%%%%%%%%%%%%%%%%%%%%%%%%%%%%%%%%%%%%%%%%%%%%%%%%%%%%%%%

%%%%%%%%%%%%%%%%%%%%%%%%%%%%%%%%%%%%%%%%%%%%%%%%%%%%%%%%%%%%%%%%%%%%%%%%%%%%%%%

\begin{widetext}
%======================================================================
\appendix*
%======================================================================

%----------------------------------------------------------------------
\section*{Appendix}
\subsection*{Recursion Relations for Bayesian Updating}
%----------------------------------------------------------------------

The angular probability distribution for $\varphi$ given the $M$ measurement outcomes $r_m=\pm1$ in directions $\alpha_m$, which are multiples of $\pi/2$, is given by
\begin{equation}
	p(\varphi\vert r_1,\dotsc,r_M) = \frac{1}{\mathcal{N}} \prod_{m=1}^{M}	\frac{ 1+r_m \cos(\varphi-\alpha_m) }{ 2 }
\end{equation}
with normalization $\mathcal{N}$, and can be expanded into a Fourier sum
\begin{equation}
	p(\varphi\vert r_1,\dotsc,r_M) = \frac{c_M(0)}{2} + \sum_{q=1}^M \Big( c_M(q) \cos(q\varphi) + s_M(q) \sin(q\varphi) \Big),
\end{equation}
where the normalization is solely contained in the coefficient $c_M(0)$.
Updating this probability distribution with the next measurement result $r_{M+1}$ amounts to multiplication with the factor $\big(1+r_{M+1}\cos(\varphi-\alpha_{M+1})\big)/2$, which we again expand into a Fourier sum. Comparing the coefficients we obtain the following recursion relations for the $c_{M+1}(q)$ and $s_{M+1}(q)$:

\begin{align}
	c_{M+1}(0) &= \frac{c_M(0)}{2} + \frac{r_{M+1}}{2} \Big(
		\cos(\alpha_{M+1}) c_M(1) +
		\sin(\alpha_{M+1}) s_M(1)
		\Big) \\
	c_{M+1}(q) &= \frac{c_M(q)}{2} + \frac{r_{M+1}}{4} \Big[
		\cos(\alpha_{M+1}) \Big( c_M(q-1) + c_M(q+1) \Big) -
		\sin(\alpha_{M+1}) \Big( s_M(q-1) - s_M(q+1) \Big)
		\Big] \\
	s_{M+1}(q) &= \frac{s_M(q)}{2} + \frac{r_{M+1}}{4} \Big[
		\sin(\alpha_{M+1}) \Big( c_M(q-1) - c_M(q+1) \Big) +
		\cos(\alpha_{M+1}) \Big( s_M(q-1) + s_M(q+1) \Big)
		\Big],
\end{align}
where for $q>M$ we set $c_M(q)=s_M(q)=0$. The starting distribution is the flat prior $p(\varphi)=1/(2\pi)$ with $c_0(0)=1/\pi$.

The advantage of the Fourier representation is that circular moments of the probability distribution can be straightforwardly calculated:
\begin{align}
\text{Normalization:} && \int_0^{2\pi} \dd{}\varphi\; p(\varphi\vert r_1,\dotsc,r_M) \phantom{\e^{i\varphi}} &= \pi c_M(0), \\
\text{first circular moment:} && \int_0^{2\pi} \dd{}\varphi\; p(\varphi\vert r_1,\dotsc,r_M) \e^{i \varphi} &= \pi \Big( c_M(1) + i s_M(1) \Big) = R.
\end{align}
The first circular moment $R$ gives rise to the mean angle $\bar{\varphi}=\arg R$ and the circular standard deviation $\sigma =\sqrt{-2\ln \abs{R}}$ \cite{Fisher,Mardia}.

\end{widetext}

%%%%%%%%%%%%%%%%%%%%%%%%%%%%%%%%%%%%%%%%%%%%%%%%%%%%%%%%%%%%%%%%%%%%%%%%%%%%%%%
\end{document}